\documentclass[10pt,oneside]{article}
\usepackage{vmargin}
\usepackage{graphicx}

\newcommand\captionof[1]{\def\@captype{#1}\caption}

\begin{document}
\title{Influence of manganese contamination on high-mobility
GaAs/AlGaAs heterostructures}
\author{K. Wagenhuber, H.-P. Tranitz, M. Reinwald, and W.
Wegscheider\\Institut f\"ur Experimentelle und Angewandte Physik,
Universit\"at Regensburg,\\ 93040 Regensburg, Germany } \maketitle

\begin{abstract}
Photoluminescence and magnetotransport measurements have been
performed to assess the quality of modulation doped GaAs/AlGaAs
heterostructures. The temporal evolution of the low-temperature
electron mobility of samples prepared in a molecular beam epitaxy
chamber containing manganese as a source material was studied. Mn
contamination was identified to be responsible for the reduction
of the electron mobility to 1~$\times$ 10$^{6}$ cm$^{2}$/Vs and
the appearance of a distinct photoluminescence band. In contrast,
structures in which this signal is absent reach mobility values of
5.4 $\times$ 10$^{6}$ cm$^{2}$/Vs. This directly proves that the
epitaxy of high-mobility electron systems and structures
containing GaMnAs layers, in principle, can be combined in one
growth chamber.
\end{abstract}
\vspace{1.5cm}
Modulation doped GaAs/AlGaAs heterostructures have
been the subject of interest for several years mainly for the
excellent transport properties of the two dimensional electron gas
(2DEG) forming at the heterojunction interface. Molecular beam
epitaxy (MBE) is the predestined technique for the growth of such
structures, because it allows abrupt doping concentrations and
sharp interfaces $^{1, 2}$. The separation of the doping layer
from the electron channel drastically increases low-temperature
electron mobilities. In the highest-mobility structures residual
impurities in the GaAs layer hosting the 2DEG were identified as
the major limiting scattering mechanism~$^{3, 4}$. Thus the
essential premise for the growth of high-quality samples is
maximum purity of the MBE system and the source materials.

Another recent application area of MBE-grown structures is the
rapidly developing field of spintronics, i. e. the coupling of
traditional properties of compound semiconductor materials with
the spin properties of ferromagnetic materials $^{5}$. GaMnAs has
emerged as a prime candidate for the investigation of these
so-called diluted semiconductors $^{6}$. However, the combination
of high-mobility electron systems with these ferromagnetic
materials, which could be employed as a spin-polarizing injector
into a 2DEG, has not been attempted so far. This is due to the
general believe that Mn-impurities, once present in the MBE
ultra-high vacuum chamber, would drastically limit the achievable
electron mobilities for this growth system.

In this letter we report the influence of Mn, which can be
unintentionally incorporated during growth, on the quality of
modulation doped GaAs/AlGaAs samples. Photoluminescence was used
to identify residues of manganese in these structures, and the
electronic properties, in particular charge carrier densities and
mobilities, were determined by magnetotransport measurements. We
find low-temperature mobilities in excess of
5~$\times$~10$^6$~cm$^2$/Vs in samples which were prepared
directly after a sample containing GaMnAs had been fabricated.
Even in a sample grown on a heavily Mn-contaminated substrate
block this value is only decreased to about
1~$\times$~10$^{6}$~cm$^2$/Vs.

The standard modulation-doped single interface (MDSI) sample
structure, schematically depicted in Figure~1, was grown by MBE on
a (001)-oriented GaAs substrate typically at 635$^{\circ}$~C. The
MBE system, which was manufactured by Veeco Instruments Inc., is
based on a Varian Gen II design and equipped with high-capacity
closed-cycle refrigerator cryogenic pumps. As a buffer layer we
grew 300 nm GaAs followed by a 1000~nm thick GaAs/AlGaAs
superlattice. This concept has been shown to significantly improve
the quality of structures placed on it, as p-type impurities
migrating from the substrate are trapped in the superlattice
$^{7}$. Onto this buffer an undoped 1000~nm thick GaAs layer and a
50~nm thick Al$_{0.33}$Ga$_{0.67}$As spacer layer were grown. The
2DEG is formed at the GaAs/AlGaAs interface and the electrons are
supplied by a Si-$\delta$-doped AlGaAs donor layer. All
investigated samples were grown with the same set of process
parameters.

\begin{figure} [t]
\begin{center}
\includegraphics[width=8cm]{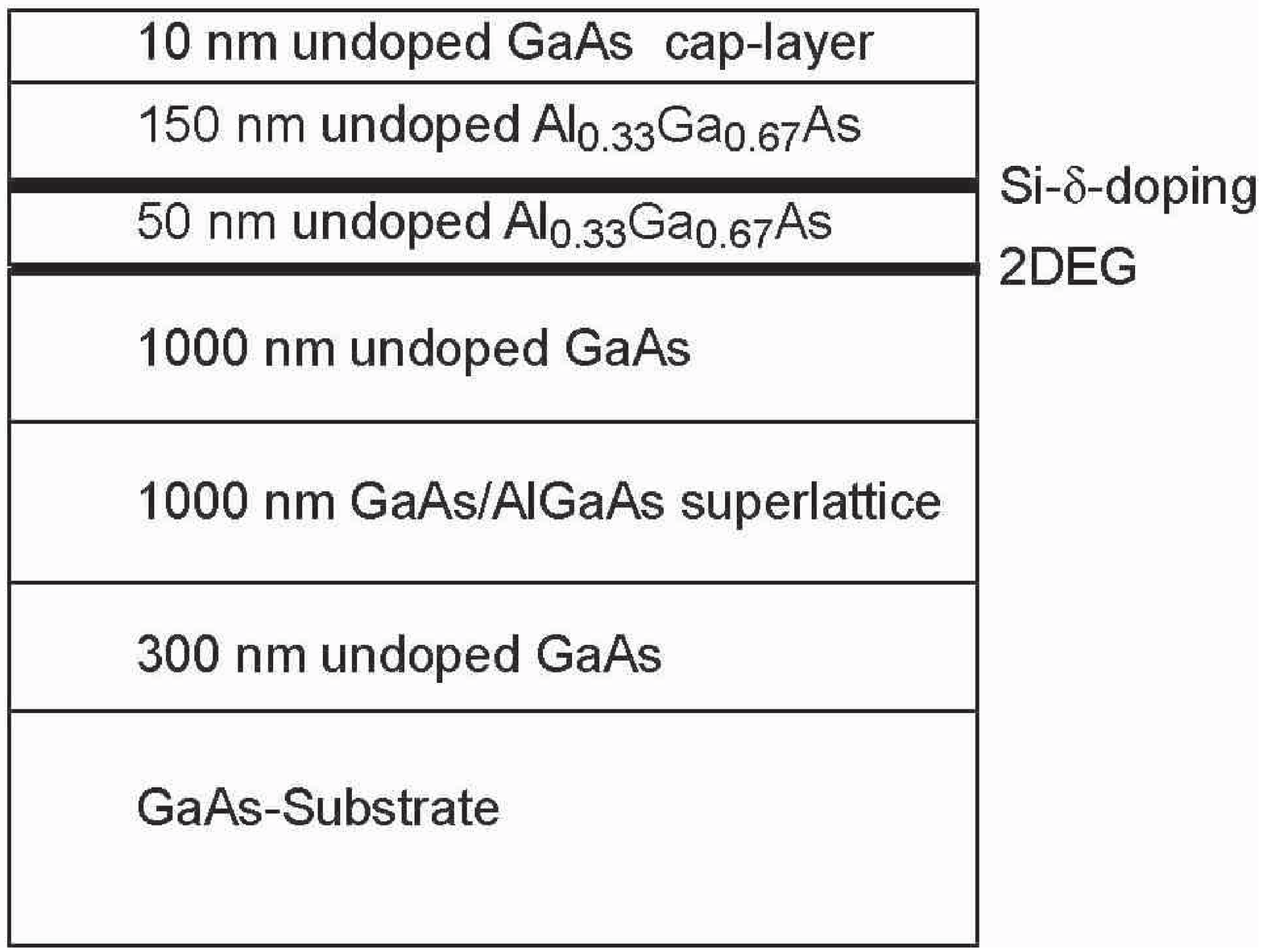}
\caption{Schematic structure of the GaAs/AlGaAs MDSI samples.}
\end{center}
\end{figure}
\begin{figure} [!b]
\begin{center}
\includegraphics[width=8cm]{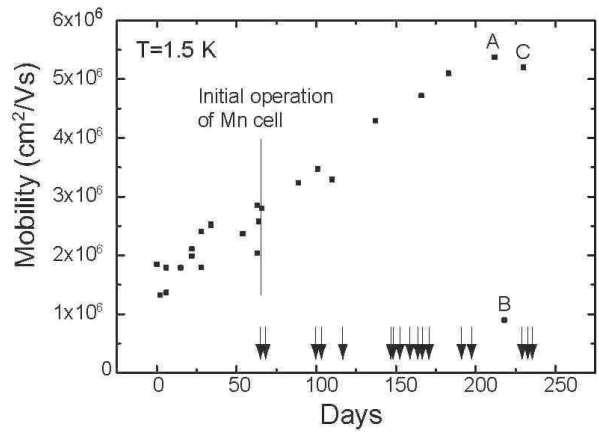}
\caption{Temporal evolution of the mobilities of our MDSI samples.
All samples except for sample B were grown on a Mn free substrate
mount. Sample A and sample C show mobilities of 5.4 and 5.2
$\times$ 10$^6$ cm$^2$/Vs respectively.}
\end{center}
\end{figure}

The electronic properties of our samples were investigated by the
Van der Pauw method at 1.5~K. In order to ionize deep DX centers
in the AlGaAs, each sample was illuminated using a red light
emitting diode before the measurement. Figure 2 illustrates the
temporal evolution of the mobilities of these standard MDSI
structures, which were grown in an MBE system equipped with a Mn
effusion cell. The vertical line indicates the initial startup of
the manganese cell and the arrows at the bottom of the graph mark
dates when GaMnAs or GaAlMnAs samples with typical Mn
concentrations of 3 to 5 \% were prepared. We find a steady
increase of mobility at consistent electron densities of about
3~$\times$~10$^{11}$~cm$^{-2}$. A maximum value in mobility of
5.4~$\times$~10$^{6}$~cm$^2$/Vs was reached after 210 days of
operation (sample A). In order to study the influence of manganese
one sample, denoted as sample B, was intentionally grown on a
heavily Mn-contaminated substrate holder. This holder was
exclusively used for GaMnAs growth so far. This led to a drop in
electron mobility down to 1.0~$\times$~10$^6$~cm$^2$/Vs. Sample C
was grown some days later on a clean mount and exhibits again a
mobility of more than 5~$\times$~10$^6$~cm$^2$/Vs. This result
makes clear that the decrease in mobility of sample~B is
definitely due to the incorporation of Mn. In addition, no memory
effect is observed. Obviously the initial startup of the Mn cell
and its regular use did not affect the quality of the samples.

In order to prove the incorporation of manganese, we performed
photoluminescence measurements of sample A and~B.
Photoluminescence is a highly sensitive tool for the detection of
residual impurities in semiconductor crystals. In addition to the
bright excitonic luminescence, defect related recombination bands
can be observed. After excitation of the sample, donor bound
electrons can recombine with acceptor bound holes. This donor
acceptor pair transitions yield a characteristic luminescence
band. The energetic position of its maximum in intensity strongly
depends on the ionization energies of the participating defects
and thus on the type of the impurities. The intensity gives
qualitative information about the concentration of the defects. In
GaAs, manganese on a gallium-site forms a deep acceptor level at
about 0.13~eV above the top of the GaAs valence band. It is
well-known that the Mn-related low-temperature PL-signal in GaAs
appears at about 1.41~eV in the spectrum. Its origin is the donor
acceptor pair transition $^{8, 9}$. This gives us a sensitive tool
to proof the presence of Mn in the GaAs layer of MDSI samples even
at very low concentrations.

\begin{figure} [t]
\begin{center}
\includegraphics[width=8cm]{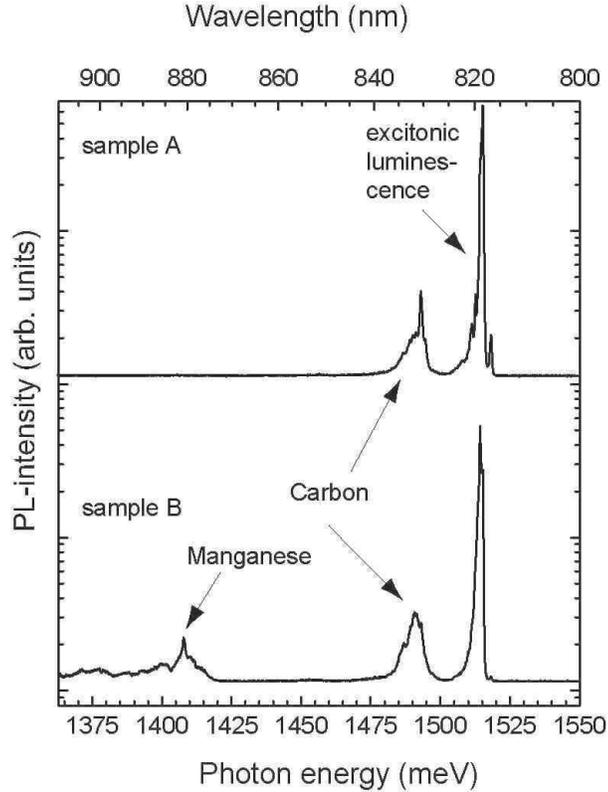}
\caption{PL spectra of sample A and sample B at 1.6 K and an
excitation density of about 0.5 mW/cm$^2$. Sample B exhibits a
Mn-related luminescence band at 1.41 eV. In sample A no Mn-related
PL signal was detected.}
\end{center}
\end{figure}

We used a He-Ne-Laser with a maximum output power of 12~mW at
$\lambda$ = 632.8 nm as an excitation source. The excitation
intensity was varied by means of neutral density filters. The
laser beam is focused on the sample (spot size d $\approx$~300
$\mu$m), which is placed in a helium bath cryostat and cooled down
to 1.6~K. For the detection of the PL emission we used a
combination of a Jobin-Yvon 0.64m monochromator and a nitrogen
cooled CCD camera. All information gained by PL is strictly
limited to the upper region of the 1000~nm thick GaAs layer for
two reasons. First, Al$_{0.33}$Ga$_{0.67}$As is not excited by
photons with an energy of 1.96~eV (${\mathrel{\widehat{=}}}$~632.8
nm) because its band gap energy is about 2~eV. Second, the
penetration depth of the 632.8~nm He-Ne laser wavelength in GaAs
is about 230~nm, which means that all of the light is absorbed by
GaAs and neither the buffer layer nor the substrate are excited
$^{10}$. Figure~3 shows the PL spectra of sample A and sample B at
excitation densities of about 0.5~mW/cm$^2$. Both are dominated by
the excitonic recombination lines of bulk GaAs at about 1.51~eV.
Further investigations of the excitonic luminescence revealed
narrow lines with full width at half maxima (FWHM) smaller than
0.25~meV, which is beyond the spectral resolution of our detection
system. This result reveals the high crystal quality and purity of
our samples. Both samples show another emission band at about 1.49
eV corresponding to the free-to-bound transition involving the
Carbon acceptor. Carbon as an unintentional impurity is well-known
but its source is still unclear. Reaction of carbon monoxide or
dioxide with either surface arsenic or gallium suboxide
(Ga$_{2}$O), and a free carbon atom is most probable $^{11}$. In
contrast to sample A, sample B exhibits an additional PL band at
about 1.41~eV. Due to the fact that sample B was grown on the
Mn-contaminated substrate holder and its energetic position we
assign this signal to the Mn impurity. Obviously, desorption and
diffusion processes of Mn, which was deposited on the substrate
holder earlier, result in an incorporation of Mn in sample B. This
Mn-related luminescence band tends to saturate in intensity by
increasing excitation density (see Figure 4). The reason for this
effect is the finite number of impurities in the material. When
all acceptors and donors are neutralized by the charge carriers,
this saturation regime is reached. In sample A, which was grown on
a clean substrate holder, no Mn-related PL signal could be
observed, even at long integration times and high excitation
densities.
\begin{figure} [t]
\begin{center}
\includegraphics[width=8cm]{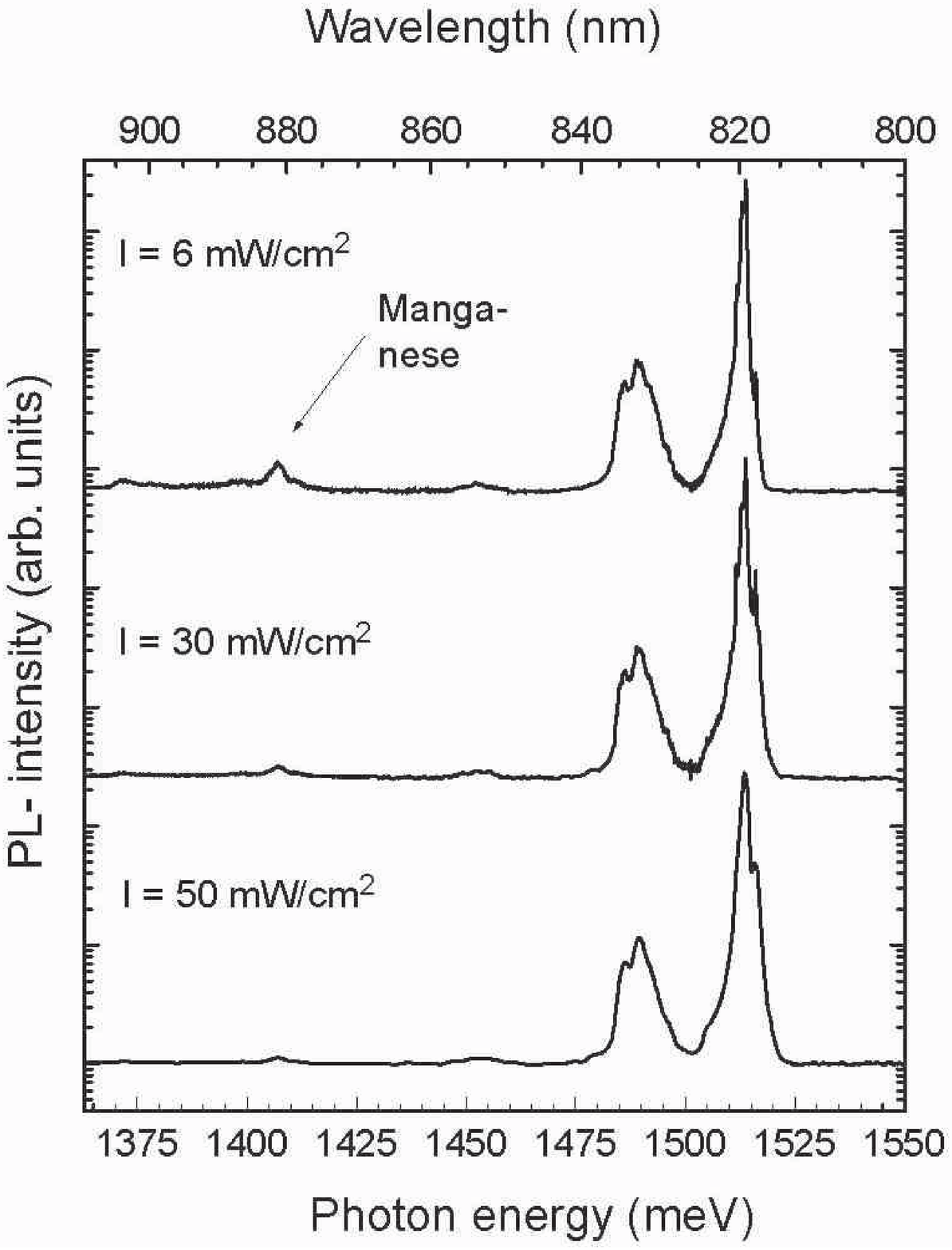}
\centering
 \caption{PL spectra of sample B at three different
excitation densities, measured at 1.6 K. All spectra are
normalized to the same excitonic intensity.}
\end{center}
\end{figure}
In summary, PL measurements of modulation doped GaAs/AlGaAs
samples which were grown in a Mn-contaminated MBE chamber have
been performed. We were able to detect a Mn-related luminescence
band in the spectrum of samples grown on a Mn-contaminated
substrate holder. Samples which were mounted on a clean substrate
holder for growth did not exhibit any Mn-related peak in the
photoluminescence signal. Therefore we conclude that these samples
are, to a high degree, free of undesired manganese impurities.
Magnetotransport measurements showed that Mn incorporation even at
rather low concentrations reduces the low-temperature electron
mobility. Nevertheless, we are able to grow high-quality MDSI
structures with mobilities in excess of 5~$\times$~10$^6$
cm$^2$/Vs in an MBE-system, which is also used for the growth of
GaMnAs samples. Most important, no memory effect was observed over
the time of
operation for GaMnAs growth of roughly half a year. \\
\\ \\ \\
\bfseries References: \\ \\ \mdseries
 $^{1}$R. Dingle, H. L. Stormer, A.
C. Gossard, and W. Wiegmann,
Appl. Phys. Lett. {\bf{33}}, 665 (1978).\\ \\
$^{2}$H. L. Stormer, R. Dingle, A. C. Gossard, W. Wiegmann, and M.
D. Sturge, Solid State Commun. {\bf{29}}, 705 (1979).\\ \\
$^{3}$V. Umansky, R. d. Piciotto, and M. Heiblum, Appl. Phys.
Lett. {\bf{71}}, 683 (1997).\\ \\
$^{4}$L. Pfeiffer, K. W. West, H. L. Stormer, and K. W. Baldwin,
Appl. Phys. Lett. {\bf{55}}, 1888 (1989).\\ \\
$^{5}$H. Ohno, Science {\bf{281}}, 951 (1998).\\ \\
$^{6}$F. Matsukura, H. Ohno, A. Shen, and Y. Sugawara, Phys. Rev.
{\bf{B}} {\bf{57}} R2037 (1998) .\\ \\
$^{7}$T. Saku, Y. Hirayama, and Y. Horikoshi, Jpn. J. Appl. Phys. {\bf{30}}, 902 (1991).\\ \\
$^{8}$M. Ilegems, R. Dingle, and L. W. Rupp Jr., J. Appl. Phys.
{\bf{46}}, 3059 (1975).\\  \\
$^{9}$W. Schairer, and M. Schmidt, Phys. Rev. {\bf{B}} {\bf{10}},
2501 (1974). \\ \\
$^{10}$D. E. Aspnes, and A. A. Studna, Phys. Rev. {\bf{B}}
{\bf{27}}, 985 (1983).\\ \\
$^{11}$M. A. Herman, and H. Sitter, Molecular Beam Epitaxy
(Springer, Berlin, 1996).\\ \\

\end{document}